\documentclass[reprint, amsmath,amssymb, aps,]{revtex4-1}

\usepackage{graphicx}
\usepackage{dcolumn}
\usepackage{bm}
\usepackage{float}
\usepackage[utf8]{inputenc}
\usepackage{longtable}
\usepackage{amsmath}
\usepackage{color}
\usepackage{amssymb}
\usepackage{hyperref}

\begin{document}

\preprint{APS/123-QED}

\title{Strain Tunable Intrinsic Ferromagnetic in 2D Square CrBr$_2$}

\author{Fei Li}
\affiliation{State Key Laboratory of Metastable Materials Science and Technology \& Key Laboratory for Microstructural Material Physics of Hebei Province, School of Science, Yanshan University, Qinhuangdao 066004, China}
\author{Yulu Ren}
\affiliation{State Key Laboratory of Metastable Materials Science and Technology \& Key Laboratory for Microstructural Material Physics of Hebei Province, School of Science, Yanshan University, Qinhuangdao 066004, China}
\author{Wenhui Wan}
\affiliation{State Key Laboratory of Metastable Materials Science and Technology \& Key Laboratory for Microstructural Material Physics of Hebei Province, School of Science, Yanshan University, Qinhuangdao 066004, China}
\author{Yong Liu}
\affiliation{State Key Laboratory of Metastable Materials Science and Technology \& Key Laboratory for Microstructural Material Physics of Hebei Province, School of Science, Yanshan University, Qinhuangdao 066004, China}
\author{Yanfeng Ge}
\email{yfge@ysu.edu.cn}
\affiliation{State Key Laboratory of Metastable Materials Science and Technology \& Key Laboratory for Microstructural Material Physics of Hebei Province, School of Science, Yanshan University, Qinhuangdao 066004, China}

\date{\today}

\begin{abstract}
Two-dimensional (2D) intrinsic magnetic materials with high Curie temperature (\emph{T}$_c$) coexisting with 100\% spin-polarization are highly desirable for realizing promising spintronic devices. In the present work, the intrinsic magnetism of monolayer square CrBr$_2$ is predicted by using first-principles calculations. The monolayer CrBr$_2$ is an intrinsic ferromagnetic (FM) half-metal with the half-metallic gap of 1.58 eV. Monte Carlo simulations based on the Heisenberg model estimates \emph{T}$_c$ as 212 K. Furthermore, the large compressive strain makes CrBr$_2$ undergo ferromagnetic-antiferromagnetic phase transition, when the biaxial tensile strain larger than 9.3\% leads to the emergence of semiconducting electronic structures. Our results show that the intrinsic half-metal with a high \emph{T}$_c$ and controllable magnetic properties endow monolayer square CrBr$_2$ a potential material for spintronic applications.
\end{abstract}

\maketitle


\section{INTRODUCTION}
Spintronic offers tremendous developments in the field of quantum computing and the next generation information technology \cite{1,2}. 100\% spin-polarization FM half-metal is a key ingredient of the high performance spintronic devices \cite{3}, one spin channel exhibits metallic property, the other spin channel has a energy gap as a semiconductor. Since the half-Heusler alloy ferromagnetic materials NiMnSb and PtMnSb are predicted in 1980s \cite{4}, there have been further researches into magnetic half-metals, such as RbSe, CsTe, NbF$_3$, CoH$_2$, ScH$_2$ and so on \cite{5,6,7}.

With the explosive research of 2D materials in the past decade, 2D FM materials have been highly investigated as one of the ideal candidates for nano-spintronic devices \cite{8}. Although FM half-metals are also discovered in 2D materials, most of them are still realized by the external conditions, such as pressure or doping \cite{9,10}. So far, the intrinsic completely spin-polarized character is relatively rare in natural 2D materials \cite{11}. Recently, as a new class of 2D materials, metal dihalides (MX$_2$, X=Cl, Br, I) are promising nanoelectronic devices due to the half-metallicity. The experiments observe that bulk metal dihalides have natural layered structure, similar to many other 2D materials \cite{12}. Among them, FeCl$_2$ is the classic experimentally known material with half-metallicity in monolayer form \cite{8}. The half-metallic gap and spin gap are about 1.0 eV and 4.4 eV, respectively, but the low $T_c$ of 17 K \cite{13} greatly blocks the prospects in spintronics. In addition, there are a flurry of researches focusing on the magnetic half-metals, such as TiCl$_3$, VCl$_3$, and MnX (X=P, As)\cite{14}. But, the intrinsic half-metallic material with wide spin gap and high $T_c$ is still absent \cite{15,16,17}. Thus, the exploration of 2D intrinsic FM half-metals is still an important frontier.

In this work, the structure of monolayer square CrBr$_2$ is identified by the particle swarm optimization (PSO) method \cite{18,19}, within an evolutionary algorithm as implemented in the CALYPSO code \cite{20}. Then, we systematically investigate the magnetic states of monolayer CrBr$_2$ by using first-principles calculations. The results show that it is an intrinsically half-metallic ferromagnetism with the wide half-metallic gap of 1.58 eV and spin gap of 3.72 eV. More strikingly, the electronic structures and magnetic configurations can be controlled by the feasible uniaxial and biaxial strains. A uniaxial compressive strain of 2.7\% or a biaxial compressive strain of 2.1\% causes a magnetic phase transition from FM to AFM. The biaxial tensile strain larger than 9.3\% can open the spin-up gap of monolayer CrBr$_2$, thus, the electronic performance transfers from half-metal into semiconductor. At last, monolayer square CrBr$_2$ is predicted to have a high $T_c$ of 212 K, which can be improved to about 404 K by the strain. Our calculations indicate that monolayer CrBr$_2$ is potentially promising for spintronic devices.

\section{METHODS}
The low-energy structure of 2D square monolayer CrBr$_2$ was identified by the PSO method \cite{18,19} within an evolutionary algorithm as implemented in the CALYPSO code \cite{20}. In the PSO simulation, the population size and the number of generations were set as 30 and 50, respectively. The monolayer CrBr$_2$ was set in the xy plane with a buckled structure along the z direction and the vacuum space was set at 20 {\AA}. The structural study in the first generation was based on random structures generated automatically using imposed symmetry constraints. At each step, only 60\% of the structures were taken into the next generation, while the other structures were generated randomly to guarantee the structural diversity. Density functional theory (DFT) calculations were performed using the projector augmented wave method, as implemented in the Vienna ab initio simulation package VASP \cite{21,22,23}. We used a Perdew-Burke-Ernzerhof (PBE) type generalized gradient approximation (GGA) in the exchange-correlation functional \cite{24}. The plane-wave cutoff energy 450 eV and Monkhorst-Pack special k-point mesh of $11\times11\times1$ for the Brillouin zone integration were used in all calculations \cite{25}. The HSE06 hybrid functional was employed to obtain the accurate electronic band gap \cite{26}. A conjugate-gradient algorithm was employed for geometry optimization using convergence criteria of $10^{-5}$ eV for the total energy and 0.01 eV/{\AA} for Hellmann-Feynman force components. The supercell size of $4 \times 4 \times 1$ unit cells was built to calculate the phonon spectrum by using the density functional perturbation theory (DFPT) \cite{27}. In order to estimate the Curie temperature, we mapped the system by Monte Carlo simulations based on the Heisenberg model in the $2 \times 2 \times 1$ supercell.

\section{RESULTS AND DISCUSSION}

\begin{figure}[t!]
\centerline{\includegraphics[width=0.4\textwidth]{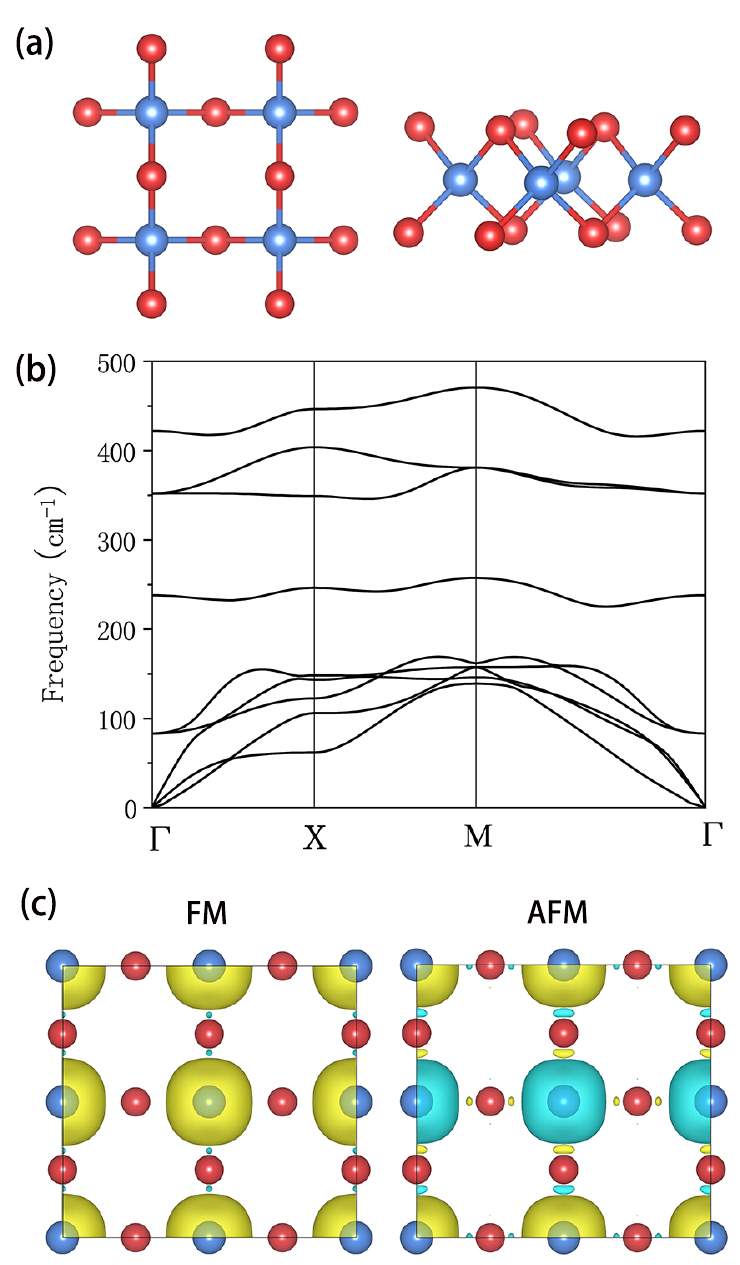}}
\caption{(a) Top and side views of monolayer square CrBr$_2$ structure. Cr and Br atoms are displayed as blue and red spheres, respectively. (b) Theoretical phonon spectrum of monolayer CrBr$_2$ obtained from DFPT calculations. (c) The spin polarization distribution for FM and AFM magnetic configurations, respectively. The yellow and blue colors indicate the net spin-up and spin-down polarization, respectively.}
\label{fig1}
\end{figure}

As shown in Fig.\ref{fig1}(a), the optimized lattice parameters for CrBr$_2$ are a=b=3.91 {\AA}. Each layer of CrBr$_2$ is composed of three atomic layers: a layer of Cr sandwiched between two layers of Br. The Cr-Br bond length is 2.60 {\AA} and the Cr-Br-Cr bond angle is about $103^{\circ}$. The absence of imaginary phonon modes in phonon spectrum indicates the dynamic stability of CrBr$_2$ [Fig.\ref{fig1}(b)].

To determine the preferred magnetic ground state, we considered two different magnetic configurations, i.e. FM configuration and AFM configuration [Fig.\ref{fig1}(c)]. The energy difference $\Delta$E ($\Delta$E= E$_{\rm AFM}$-E$_{\rm FM}$) of 0.176 eV in $2 \times 2 \times 1$ supercell indicates an FM ground state of CrBr$_2$. Additionally, the energy difference of 6.54 eV between the non-magnetic state and the magnetic state is too huge to neglect the nonmagnetic state \cite{15}. CrBr$_2$ has FM ordering with large magnetic moments and the magnetism mainly stems from the Cr atom when Br atom holds small opposite spin moments [Fig.\ref{fig1}(c)]. The square crystal structure renders the magnetic state governed by significant competition between two mechanisms \cite{29}. One is the direct AFM interaction between two neighboring Cr atoms. The other is superexchange FM interaction of two neighboring Cr atoms mediated by Br atom. The preference for either FM or AFM ordering can be rationalized using Goodenough-Kanamori-Anderson (GKA) formalism \cite{30,31,32}. In monolayer CrBr$_2$, Cr-Br-Cr bond angle is ~$103^{\circ}$ closer to ~$90^{\circ}$, which usually associates with FM ordering according to GKA rules. So the neighboring Cr atoms favor superexchange to direct interaction, which leads to FM ground state.

\begin{figure*}[t!]
\centerline{\includegraphics[width=\textwidth]{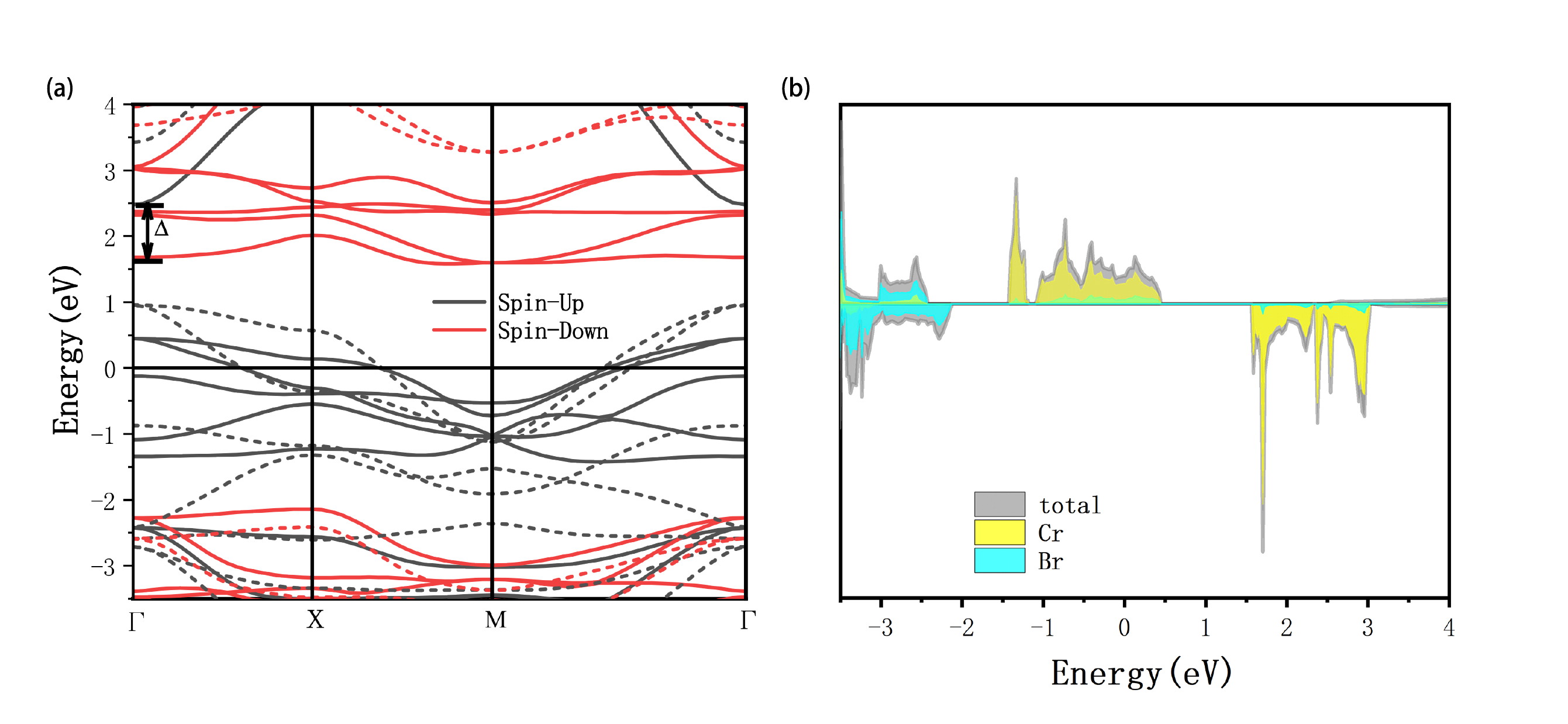}}
\caption{(a) Band structure of monolayer CrBr$_2$ with PBE functional (solid line) and HSE06 hybrid functional (dotted line). The spin-up and spin-down bands are displayed as black and red lines, respectively. (b)The projected densities of states of monolayer CrBr$_2$.}
\label{fig2}
\end{figure*}

\begin{figure*}[t!]
\centerline{\includegraphics[width=\textwidth]{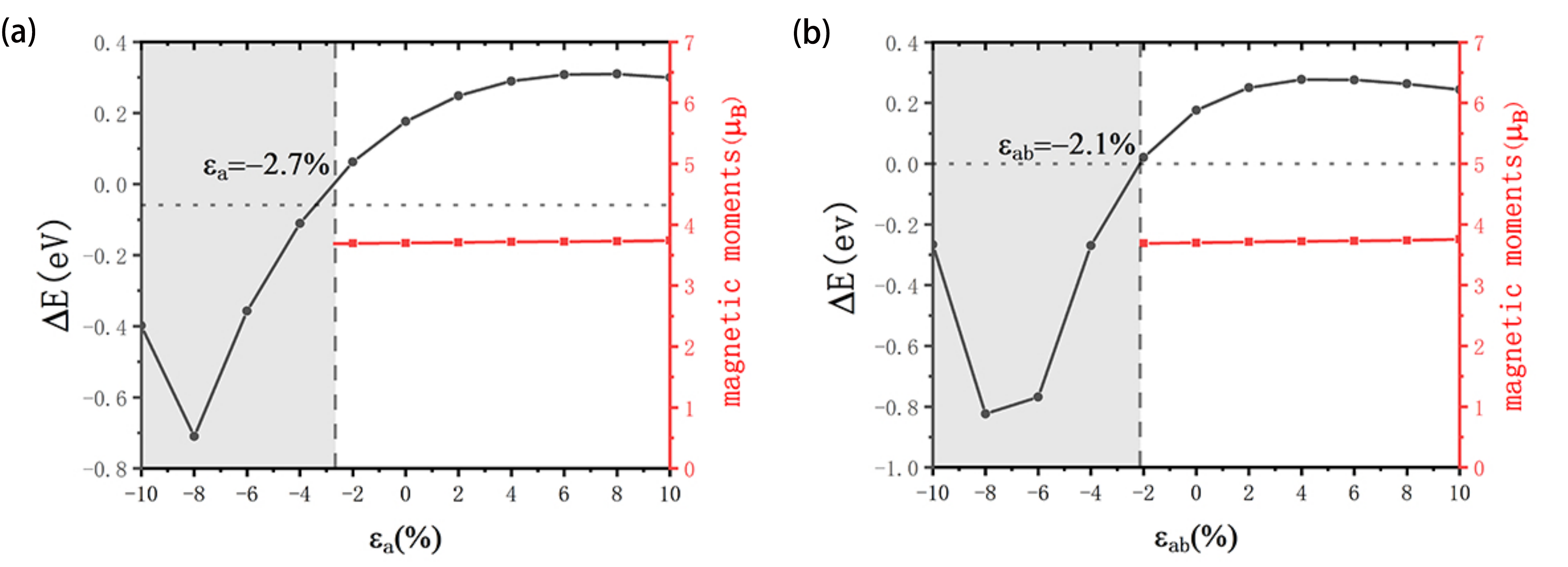}}
\caption{(a, b) Energy difference between the AFM and FM phases and the projected magnetic moments of each Cr atom (red line) of monolayer CrBr$_2$ under various uniaxial (a) and biaxial (b) strain. The projected magnetic moments of each Cr atom is hardly affected when the strain range from -2\% to 10\%.}
\label{fig3}
\end{figure*}

Notably, the electronic structures of monolayer CrBr$_2$ with PBE and HSE06 functionals show the spin-up states cross the Fermi level, while the spin-down channel acts as a semiconductor [Fig.\ref{fig2}]. Furthermore, the intrinsic half-metallic property of monolayer CrBr$_2$ is also ensured by the GGA+U functional \cite{44}[Appendix A]. As a key parameter, the half-metallic gap is defined as the minimum of the difference between Fermi level and the bottom of spin-down conduction bands, and the difference between Fermi level and the top of spin-down valence bands \cite{12}. With the PBE functional, the half-metallic gap is 1.58 eV, which raises to 2.41 eV with HSE06 functional. It is large enough to efficiently prevent the thermally agitated spin-flip transition. The difference between the bottom of spin-down conduction bands and the top of spin-down valence bands in the half-metal is defined as spin-gap, which is as high as 3.72 (5.69) eV with PBE (HSE06) functional. In addition, the large spin exchange splitting is crucial for the spin-polarized carrier injection and detection. The spin exchange splitting is 0.93 eV [labeled as $\Delta$ in Fig.\ref{fig2}(a)] for monolayer CrBr$_2$, larger than 0.24 eV in CrGeTe$_3$ \cite{33}.

\begin{figure*}[t!]
\centerline{\includegraphics[width=\textwidth]{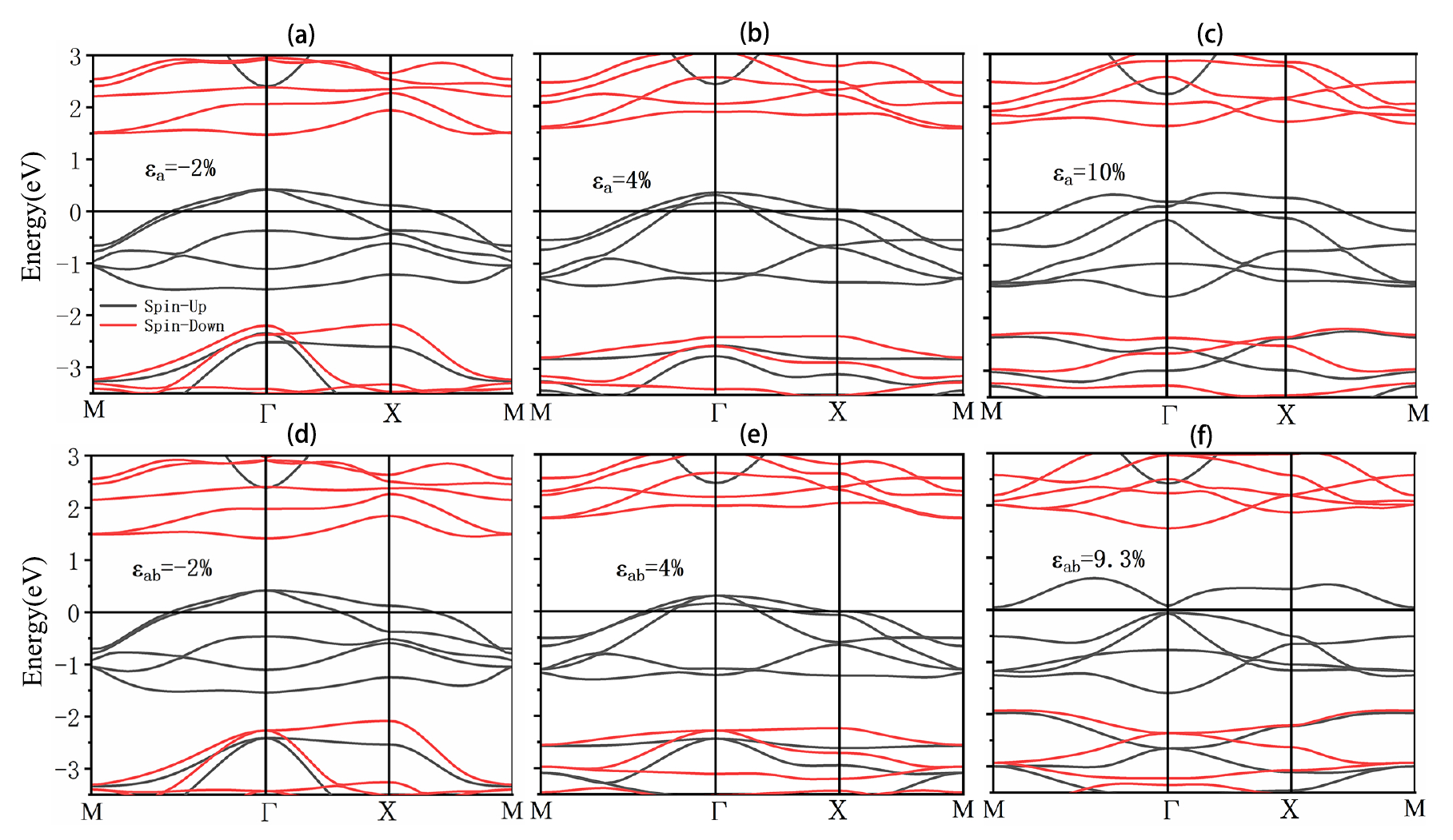}}
\caption{(a-c) Band structures of CrBr$_2$ under uniaxial strain from -2\%, 4\% and 10\%, respectively. (d-f) The band structures of monolayer CrBr$_2$ under biaxial strain of -2\% , 4\% and 9.3\%, respectively.}
\label{fig4}
\end{figure*}

As a common method in experiments, strain engineering is a well-controlled scheme to tune the electronic and magnetic properties of 2D materials in previous works \cite{34,35,36,37,38,39}. Here, we also calculate the electronic and magnetic properties of monolayer CrBr$_2$ under uniaxial and biaxial strain range from -10\% to 10\%. Figure.\ref{fig3} shows the energy difference between the FM and AFM phases and the projected magnetic moments of each Cr atom monolayer CrBr$_2$ under various uniaxial and biaxial strains. The change in $\Delta$E indicates a possible FM-AFM phase transition occurs around the uniaxial compressive strain of 2.7\%, or the biaxial compressive strain of 2.1\%. In the band structures under uniaxial and biaxial strain [Fig.\ref{fig4}], the half-metallic gap becomes smaller (larger) under compressive (tensile) strain. It is more noteworthy that the opened spin-up band gap shows the characteristics of a semiconductor under the biaxial tensile strain within the reasonable range of 9.3\%-10\%.

\begin{table}[H]
\caption{\label{a} The \emph{T}$_c$ (K), -ICOHP (Cr-Br) and -ICOHP (Cr-Cr) for the monolayer CrBr$_2$ under uniaxial strain.}
\begin{tabular}{ccccc}
\toprule
\textbf{System}	& \textbf{$\varepsilon_a$=2\%}	& \textbf{$\varepsilon_a$=4\% } & \textbf{$\varepsilon_a$=6\% } & \textbf{$\varepsilon_a$=8\% } \\
$T_c$		& 310                 & 388                 & 400                 & 404 \\
-ICOHP(Cr-Br)& 2.29                & 2.31                & 2.33                & 2.33 \\
-ICOHP(Cr-Cr) & 0.31                & 0.27                & 0.25                & 0.23 \\
\label{tab}
\end{tabular}
\end{table}

Finally, the present work employs Monte Carlo simulations with the Heisenberg model to predict $T_c$. The spin Hamiltonian of monolayer CrBr$_2$ can be considered as ${H =  - \sum\limits_{ < ij > } {{J_{ij}}{S_i}{S_j}}-\sum\limits_{ i }A(S_{i}^{z})^{2}}$, where $\emph{J}_{ij}$ represents the exchange interaction parameter of the nearest neighbor Cr-Cr pairs, $S_i$ represents the spin of atom i, A is magnetocrystalline anisotropy energy parameter (MAE), and $S_{i}^{z}$ is the spin component along the z direction. The exchange interaction parameters $\emph{J}_{ij}$ are determined by expressing the total energy of the FM and AFM configurations. With $\emph{J}_{ij}$ of 11.0 meV and MAE of 126 $\mu$eV, \emph{T}$_c$ of monolayer CrBr$_2$ is about 212 K and is much higher than those in the 2D FM half-metals reported early, e.g. monolayer FeCl$_2$ (17 K) \cite{13}. Besides, the negative integrated crystal orbital Hamilton population (-ICOHP)\cite{40,41,42,43} and \emph{T}$_c$ of CrBr$_2$ under different strain are summarized in Tab.\ref{tab}. As mentioned above, FM ground state is driven by direct versus indirect exchange interaction between two neighboring Cr atoms. The results show that the direct interaction (Cr-Cr) is reduced, but the superexchange (Cr-Br) is enhanced, which leads to the enhanced FM and the increased $T_c$ in CrBr$_2$. \emph{T}$_c$ rises to a maximum of nearly 404 K under the strain of 8\%. Moreover, under the large biaxial tensile strain ($>$9.3\%), CrBr$_2$ is a room-temperature FM semiconductor with \emph{T}$_c$ of 297 K.

\section{Conclusion}
In conclusion, we have made a systematic computation of stable monolayer square CrBr$_2$ by using first-principles calculations. The electronic structures show that CrBr$_2$ exhibits ferromagnetic half-metal properties with a large half-metallic gap of 1.58 eV and the spin gap of 3.72 eV at the PBE functional level. In particular, the electronic and magnetic performances can be significantly modulated by the strain. Application of a uniaxial compressive strain of 2.7\% or a biaxial compressive strain of 2.1\% causes the FM-AFM transition. The biaxial tensile strain larger than 9.3\% results in the electronic performance transition from half-metal into semiconductor. By using Monte Carlo simulation with Heisenberg model, \emph{T}$_c$ of monolayer CrBr$_2$ is predicted to be 212 K, which can rise to 404 K after the application of strain. The intrinsic half-metallic with high \emph{T}$_c$ and controllable magnetic properties endows monolayer CrBr$_2$ a potential functional material for spintronic applications.

\begin{acknowledgments}
This work was supported by National Natural Science Foundation of China (No.11904312 and 11904313), the Project of Hebei Education Department, China (No.ZD2018015 and QN2018012), and the Natural Science Foundation of Hebei Province (No.A2019203507). Thanks to the High Performance Computing Center of Yanshan University.
\end{acknowledgments}

\section*{Appendix A}

\begin{figure}[t!]
\centerline{\includegraphics[width=0.4\textwidth]{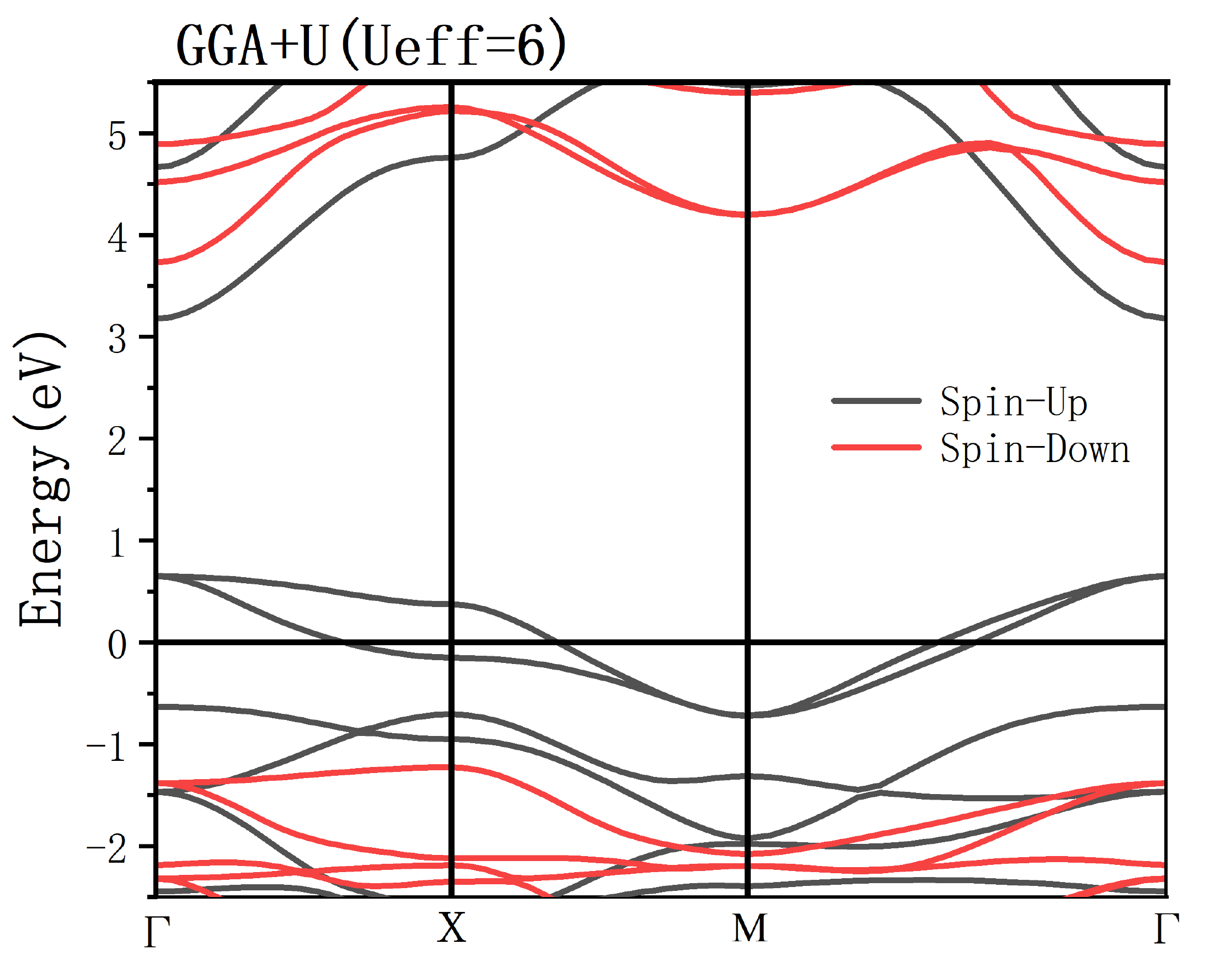}}
\caption{Electronic band structures for monolayer CrBr$_2$ with GGA+U functional.}
\label{figa}
\end{figure}


\begin{thebibliography}{0}%
\makeatletter
\providecommand \@ifxundefined [1]{%
 \@ifx{#1\undefined}
}%
\providecommand \@ifnum [1]{%
 \ifnum #1\expandafter \@firstoftwo
 \else \expandafter \@secondoftwo
 \fi
}%
\providecommand \@ifx [1]{%
 \ifx #1\expandafter \@firstoftwo
 \else \expandafter \@secondoftwo
 \fi
}%
\providecommand \natexlab [1]{#1}%
\providecommand \enquote  [1]{``#1''}%
\providecommand \bibnamefont  [1]{#1}%
\providecommand \bibfnamefont [1]{#1}%
\providecommand \citenamefont [1]{#1}%
\providecommand \href@noop [0]{\@secondoftwo}%
\providecommand \href [0]{\begingroup \@sanitize@url \@href}%
\providecommand \@href[1]{\@@startlink{#1}\@@href}%
\providecommand \@@href[1]{\endgroup#1\@@endlink}%
\providecommand \@sanitize@url [0]{\catcode `\\12\catcode `\$12\catcode
  `\&12\catcode `\#12\catcode `\^12\catcode `\_12\catcode `\%12\relax}%
\providecommand \@@startlink[1]{}%
\providecommand \@@endlink[0]{}%
\providecommand \url  [0]{\begingroup\@sanitize@url \@url }%
\providecommand \@url [1]{\endgroup\@href {#1}{\urlprefix }}%
\providecommand \urlprefix  [0]{URL }%
\providecommand \Eprint [0]{\href }%
\providecommand \doibase [0]{http://dx.doi.org/}%
\providecommand \selectlanguage [0]{\@gobble}%
\providecommand \bibinfo  [0]{\@secondoftwo}%
\providecommand \bibfield  [0]{\@secondoftwo}%
\providecommand \translation [1]{[#1]}%
\providecommand \BibitemOpen [0]{}%
\providecommand \bibitemStop [0]{}%
\providecommand \bibitemNoStop [0]{.\EOS\space}%
\providecommand \EOS [0]{\spacefactor3000\relax}%
\providecommand \BibitemShut  [1]{\csname bibitem#1\endcsname}%
\let\auto@bib@innerbib\@empty
\end{thebibliography}%


\begin{thebibliography}{100}
\expandafter\ifx\csname url\endcsname\relax
  \def\url#1{\texttt{#1}}\fi
\expandafter\ifx\csname urlprefix\endcsname\relax\def\urlprefix{URL }\fi
\providecommand{\bibinfo}[2]{#2}


\bibitem{1} Li X and Yang J \href{https://doi.org/10.1093/nsr/nww026}{2016 \emph{Natl. Sci. Rev.} \textbf{3} 365}
\bibitem{2} Feng Y P, Shen L, Yang M, Wang A, Zeng M, Wu Q, Chintalapati S and Chang C-R \href{ https://doi.org/10.1002/wcms.1313}{2017 \emph{Wires. Comput. Mol. Sci.} \textbf{7} e1313}
\bibitem{3} Felser C, Fecher G H F, and Balke B \href{https://doi.org/10.1002/anie.200601815}{2007 \emph{Angew. Chem. Int. Edi.} \textbf{46} 668}
\bibitem{4} Groot R A de, Mueller F M, Engen P G v and Buschow K H J \href{https://doi.org/10.1103/PhysRevLett.50.2024}{1983 \emph{Phys. Rev. Lett.} \textbf{50} 2024}
\bibitem{5} Xie H-H, Ma R-Y, Gao Q, Li L and Deng J-B \href{http://dx.doi.org/ https://doi.org/10.1016/j.cplett.2016.08.063}{2016 \emph{Chem. Phys. Lett.} \textbf{661} 89}
\bibitem{6} Yang B, Wang J, Liu X and Zhao M \href{https://doi.org/10.1039/C7CP04985D}{2018 \emph{Phys. Chem. Chem. Phys.} \textbf{20} 4781}
\bibitem{7} Wu Q, Zhang Y, Zhou Q, Wang J and Zeng X C \href{https://doi.org/10.1021/acs.jpclett.8b01976}{2018 \emph{J. Phys. Chem. Lett.} \textbf{9} 4260}
\bibitem{8} Ashton M, Gluhovic, Sinnott S B, Jing G and Hennig R G \href{http://dx.doi.org/ 10.1021/acs.nanolett.7b01367}{2017 \emph{Nano. Lett.} \textbf{17} 5251}
\bibitem{9} Dudarev S L, Botton G A, Savrasov S Y, Humphreys C J and Sutton A P \href{http://dx.doi.org/ 10.1103/PhysRevB.57.1505}{1998 \emph{Phys. Rev. B} \textbf{57} 1505}
\bibitem{10} Heyd J, Scuseria G E and Ernzerhof M \href{http://dx.doi.org/10.1063/1.1564060}{2003 \emph{J. Chem. Phys.} \textbf{118} 8207}
\bibitem{11} Xu R, Zou X, Liu B and Cheng H-M \href{http://dx.doi.org/ https://doi.org/10.1016/j.mattod.2018.03.003}{2018 \emph{Mater. Today} \textbf{21} 391}
\bibitem{12} Shishidou T, Freeman A J and Asahi R \href{http://dx.doi.org/10.1103/PhysRevB.64.180401}{2001 \emph{Phys. Rev. B} \textbf{64} 180401}
\bibitem{13} Torun E, Sahin H, Singh S K and Peeters F M \href{http://dx.doi.org/10.1063/1.4921096}{2015 \emph{Appl. Phys. Lett.} \textbf{106} 192404}
\bibitem{14} Zhou Y, Lu H, Zu X and Gao F \href{http://dx.doi.org/10.1038/srep19407}{2016 \emph{Sci. Rep.} \textbf{6} 19407}
\bibitem{15} Hu T, Wan W, Ge Y and Liu Y \href{http://dx.doi.org/ 10.1088/1361-648x/ab95cc}{2020 \emph{J. Phys. Condens. Matter} \textbf{32} 385803}
\bibitem{16} Gao G, Hu L, Yao K, Luo B and Liu N \href{http://dx.doi.org/ https://doi.org/10.1016/j.jallcom.2012.11.077}{2013 \emph{J. Alloy. Compd.} \textbf{551} 539}
\bibitem{17} Kobayashi K-I, Kimura T, Sawada H, Terakura K and Tokura Y \href{http://dx.doi.org/ 10.1038/27167}{1998 \emph{Nature} \textbf{395} 667}
\bibitem{18} Wang X, Li J, Botana J, Zhang M, Zhu H, Chen L, Liu H, Cui T and Miao M \href{http://dx.doi.org/ 10.1063/1.4826636}{2013 \emph{J. Chem. Phys.} \textbf{139} 164710}
\bibitem{19} Wang Y, Miao M, Lv J, Zhu L, Yin K, Liu H and Ma Y \href{http://dx.doi.org/10.1063/1.4769731}{2012 \emph{J. Chem. Phys.} \textbf{137} 224108}
\bibitem{20} Wang Y, Lv J, Zhu L and Ma Y \href{http://dx.doi.org/ https://doi.org/10.1016/j.cpc.2012.05.008}{2012 \emph{Comput. Phys. Commun.} \textbf{183} 2063}
\bibitem{21} Bl\"ochl P E \href{http://dx.doi.org/10.1103/PhysRevB.50.17953}{1994 \emph{Phys. Rev. B} \textbf{50} 17953}
\bibitem{22} Kresse G and Furthm\"uller J \href{http://dx.doi.org/10.1103/PhysRevB.54.11169}{1996 \emph{Phys. Rev. B} \textbf{54} 11169}

\bibitem{23} Kresse G and Furthm\"uller J \href{http://dx.doi.org/ https://doi.org/10.1016/0927-0256(96)00008-0}{1996 \emph{Comput. Mater. Sci.} \textbf{6} 15}
\bibitem{24} Perdew J P, Burke K and Ernzerhof M \href{http://dx.doi.org/10.1103/PhysRevLett.77.3865}{1996 \emph{Phys. Rev. Lett.} \textbf{77} 3865}
\bibitem{25} Monkhorst H J, Pack J D \href{http://dx.doi.org/10.1103/PhysRevB.13.5188}{1976 \emph{Phys. Rev. B} \textbf{13} 5188}

\bibitem{26} Heyd J, Scuseria G E and Ernzerhof M \href{http://dx.doi.org/10.1063/1.1564060}{2003 \emph{J. Chem. Phys.} \textbf{118} 8207}
\bibitem{27} Baroni S, de Gironcoli S, Corso A Dal and Gian-nozzi P \href{http://dx.doi.org/10.1103/RevModPhys.73.515}{2001 \emph{Rev. Mod. Phys.} \textbf{73} 515}

\bibitem{29} Anderson P W \href{http://dx.doi.org/10.1103/PhysRev.115.2}{1959 \emph{Phys. Rev.} \textbf{115} 2}
\bibitem{30} Goodenough J B \href{http://dx.doi.org/10.1103/PhysRev.100.564}{1955 \emph{Phys. Rev.} \textbf{100} 564}
\bibitem{31} Goodenough J B \href{http://dx.doi.org/ https://doi.org/10.1016/0022-3697(58)90107-0}{1958 \emph{J. Phys. Chem. Solids} \textbf{6} 287}
\bibitem{32} Kanamori J \href{http://dx.doi.org/10.1063/1.1984590}{1960 \emph{J. Appl. Phys.} \textbf{31} S14}
\bibitem{33} Li X and Yang J \href{http://dx.doi.org/10.1039/C4TC01193G}{2014 \emph{J. Mater. Chem. C} \textbf{2} 7071}

\bibitem{34} Yu M, Liu X and Guo W \href{http://dx.doi.org/ 10.1039/C7CP07912E}{2018 \emph{Phys. Chem. Chem. Phys.} \textbf{20} 6374}
\bibitem{35} Gong S, Wan W, Guan S, Tai B, Liu C, Fu B, Yang S A and Yao Y \href{http://dx.doi.org/ 10.1039/C7TC01399J}{2017 \emph{J. Mater. Chem. C} \textbf{5} 8424}
\bibitem{36} Wang Y and Ding Y \href{http://dx.doi.org/10.1039/C7TC05717B}{2018 \emph{J. Mater. Chem. C} \textbf{6} 2245}
\bibitem{37} Peng C, Wang Y, Cheng Z, Zhang G, Wang C and Yang G \href{http://dx.doi.org/10.1039/C5CP00364D}{2015 \emph{Phys. Chem. Chem. Phys.} \textbf{17} 16536}
\bibitem{38} Zhou W, Liu L and Wu P \href{http://dx.doi.org/ https://doi.org/10.1016/j.physleta.2014.01.021}{2014 \emph{Phys. Lett. A} \textbf{378} 909}

\bibitem{39} Ma Y, Dai Y, Guo M, Niu C, Zhu Y and Huang B \href{http://dx.doi.org/10.1021/nn204667z}{2012 \emph{ACS Nano} \textbf{6} 1695}
\bibitem{40} Dronskowski R and Bloechl P E \href{http://dx.doi.org/10.1021/j100135a014}{1993 \emph{J. Chem. Phys.} \textbf{97} 8617}
\bibitem{41} Deringer V L, Tchougr\'eeff A L and Dronskowski R \href{http://dx.doi.org/10.1021/jp202489s}{2011 \emph{J. Phys. Chem. A} \textbf{115} 5461}
\bibitem{42} Maintz S, Deringer V L, Tchougr\'eeff A L and Dronskowski R \href{http://dx.doi.org/https://doi.org/10.1002/jcc.23424}{2013 \emph{J. Comput. Chem.} \textbf{34} 2557}
\bibitem{43} Maintz S, Deringer V L, Tchougr\'eeff A L and Dronskowski R \href{http://dx.doi.org/https://doi.org/10.1002/jcc.24300}{2016 \emph{J. Comput. Chem.} \textbf{37} 1030}
\bibitem{44} Liechtenstein, A. I. and Anisimov, V. I. and Zaanen, J.article \href{https://link.aps.org/doi/10.1103/PhysRevB.52.R5467}{1995 \emph{Phys. Rev. B} \textbf{52} R5467}



\end{thebibliography}
\end{document}